\documentclass[prd,preprint,tightenlines,superscriptaddress,floatfix,showpacs,preprintnumbers,nofootinbib,eqsecnum]{revtex4}

 \usepackage[dvips,final]{graphicx}
   \usepackage{amsmath}
     \usepackage{epsfig}
      \usepackage{bm}

\begin{document}

\thispagestyle{empty} \preprint{\hbox{}} \vspace*{-10mm}

\title{Charge splitting of directed flow and
space-time picture of pion emission from 
the 
electromagnetic interactions with spectators}

\author{A.~Rybicki}
\email{andrzej.rybicki@ifj.edu.pl}

\affiliation{H.Niewodnicza\'nski Institute of Nuclear Physics, Polish Academy of Sciences, Radzikowskiego 152, 31-342 Krak\'ow, Poland}

\author{A.~Szczurek}
\email{antoni.szczurek@ifj.edu.pl}

\affiliation{H.Niewodnicza\'nski Institute of Nuclear Physics, Polish Academy of Sciences, Radzikowskiego 152, 31-342 Krak\'ow, Poland}
\affiliation{University of Rzesz\'ow, Rejtana 16, 35-959 Rzesz\'ow, 
Poland}

\date{\today}

\begin{abstract}

We estimate the effect of the spectator-induced electromagnetic 
interaction on the directed flow of charged pions. For intermediate 
centrality Au+Au collisions at $\sqrt{s_{NN}}=7.7$~GeV, we demonstrate 
that the electromagnetic interaction between spectator charges and final 
state pions results in 
 charge splitting of positive and negative pion directed flow. Such a 
charge splitting is visible in the experimental data reported by the 
STAR Collaboration.

The magnitude of this charge splitting appears to strongly 
depend on the actual distance between the pion emission site (pion at 
freeze-out) and the spectator system. As such, the above 
electromagnetic effect brings new, independent information on the 
space-time evolution of pion production in heavy ion collisions.

From the comparison of our present analysis to our earlier studies made 
for pions produced at higher rapidity, we formulate conclusions on the 
rapidity dependence of the distance between the pion emission site and 
the spectator system. This distance appears to decrease with increasing 
pion rapidity, reflecting the longitudinal expansion of the 
strongly-interacting system responsible for pion emission.
 Thus for the first time, information on the
space-time characteristics of the 
system
 is being provided {by
 means of
 the spectator-induced electromagnetic interaction}.

 The above electromagnetic effect being in fact a straight-forward 
consequence of the presence of spectator charges in the collision, we 
consider that it should be considered as a baseline for studies of other 
phenomena, like those related to the electric conductivity of the quark-gluon 
plasma.
 
\end{abstract}


\maketitle

\section{Introduction}
\label{secone}

Electromagnetic properties of heavy ion collisions, and in particular phenomena related to 
the presence of strong EM fields in the course of the nucleus-nucleus reaction, attract an 
evident interest in the community.
 While a large number of inspiring papers could be cited here, only two 
will be addressed below, as two different but highly interesting examples.
Firstly, a proposal for the estimation of the electric conductivity of the quark gluon 
plasma, by using the electric field resulting from the charge difference between the 
colliding Cu and Au nuclei was made by Hirono, Hongo and Hirano~\cite{hirono2012}. Secondly, a study of electric currents acting in the QGP, and induced by 
the magnetic fields present in Au+Au and Pb+Pb reactions, recently became 
available from G\"{u}rsoy, Kharzeev and Rajagopal~\cite{kharzeev2014}. In both cases, the 
sensitivity of azimuthal asymmetries in particle emission, and in particular directed
 flow\footnote{Note: $\phi$ denotes the emitted particle's azimuthal angle while
$\Psi_\mathrm{RP}$ gives the orientation of the reaction plane.}
$v_1\equiv \langle \cos (\phi - \Psi_\mathrm{RP}) \rangle$, to electromagnetic phenomena acting in 
heavy ion collisions was pointed 
out
by the authors. 

This corroborates with the results of our recent work
 on peripheral Pb+Pb collisions~\cite{Rybicki-v1}, where the influence of the electromagnetic interaction between 
the spectator systems and the charged final state pions was quantified. Our work 
demonstrated that this interaction could induce sizeable values of directed flow, and 
could be used as a new source of information on the space-time evolution of 
 pion production. This 
 finding
 was
 quite similar to what we established 
 earlier for 
$\pi^+/\pi^-$ ratios
at high values of rapidity~\cite{Rybicki-coulomb}.

Electromagnetic effects on particle emission clearly imply the dependence of specific 
observables on particle charge, in particular also for particles of the same mass (like 
e.g. $\pi^+$ and $\pi^-$ mesons). Such a charge dependence for specific components of pion 
directed flow was indeed predicted in~\cite{hirono2012,kharzeev2014}, as well as
 by us 
 in~\cite{Rybicki-v1}. In this context, the importance of experimental data on directed flow 
measured {\em separately} for positive and negative pion charges ($v_1^{\pi^+}$, $v_1^{\pi^-}$) 
becomes clearly evident. Such data is still, at the present moment, rather scarce.


The STAR Collaboration~\cite{starwhite} is, to the 
best of our knowledge, the first and unique experimental group in high energy heavy ion physics to 
provide such data simultaneously for positively and negatively charged pions\footnote{We
 note the existence of the data on directed and elliptic flow of (exclusively) positive pions near 
target rapidity measured by the WA98 experiment~\cite{wa98}, many years prior to the data from
 the STAR Collaboration. We also note the presence of STAR measurements on elliptic flow~\cite{Star-v2}; this was
 also made separately for particles of different charges including $\pi^+$ and $\pi^-$.}. 
 Final data of the STAR Collaboration on the
 directed flow of protons, antiprotons and pions
in 
the Au+Au collision energy range from $\sqrt{s_{NN}}=7.7$ up to 200~GeV 
 have
 recently been published~\cite{star2014}.
 These include 
 measurements of $v_1$ for $p$, $\bar{p}$, $\pi^+$ 
and $\pi^-$, made 
 altogether for seven
 collision energies in the c.m.s. rapidity range of 
$|y|<1$. Specifically, at lower values of $\sqrt{s_{NN}}$ the comparison of positive and negative pion directed flow in 
intermediate centrality (10-40\%) Au+Au collisions displays a 
 splitting 
 of 
 $v_1^{\pi^+}$ and 
$v_1^{\pi^-}$, with $v_1^{\pi^+}<v_1^{\pi^-}$ at positive rapidity. As remarked both in~\cite{kharzeev2014}
 and 
 by us 
 in~\cite{Rybicki-v1}, this is consistent with the expectation of a specific 
charge-dependent component of directed flow, induced by electromagnetic effects. 
 While already in~\cite{Rybicki-v1}, we stated our idea 
 of this effect being
 caused 
 by the
 spectator-induced
 EM 
 field,
 we feel that a 
more in-depth verification of this hypothesis is in place. It is indeed of importance to 
verify whether the electromagnetic interaction between spectator charges and {\em final 
state} pions can have the right magnitude to explain the observed splitting of $v_1^{\pi^+}$ 
and $v_1^{\pi^-}$, and, if this is indeed the case, to establish what conclusions can be 
drawn on that basis for the space-time evolution of the collision dynamics. This is even 
more important in view of the predictions made in~\cite{kharzeev2014} of the link between charged 
currents in the quark gluon plasma and the charge splitting of directed flow. The relatively 
straight-forward electromagnetic effect on final state pions studied by us can indeed be 
considered as a {\em baseline} for these more untrivial phenomena related to the electric 
conductivity of the QGP.

In this context, the aim of the present paper is to estimate the effect of the 
spectator-induced electromagnetic field on the directed flow of charged final state pions. 
The study presented here is made for the specific case of pions produced in the c.m.s. 
rapidity range $|y|<1$ corresponding to the STAR data~\cite{star2014}, in intermediate centrality 
Au+Au collisions. We focus on the lowest STAR energy of $\sqrt{s_{NN}}=7.7$~GeV where the 
effect of charge splitting of directed flow is found to be largest.

The remainder of this paper is organized as follows. The discussion of the charge splitting 
 apparent
 in STAR data is made in section~\ref{stardata}. The description of our numerical Monte 
Carlo tool, used to estimate the electromagnetic component of directed flow, is presented in 
section~\ref{model}. The results of our Monte Carlo simulation, as well as their comparison 
to the STAR data, are discussed in section~\ref{results}. 
 In section~\ref{implications}, we compare our results with 
these coming from our earlier 
studies,
 and comment on other effects that could induce charge-dependent pion directed flow.
 Our conclusions are summarized in section~\ref{conclusions}.


\section{Experimental data from STAR}
\label{stardata}

\begin{figure}[t]             
\centering
\includegraphics[width=0.5\textwidth]{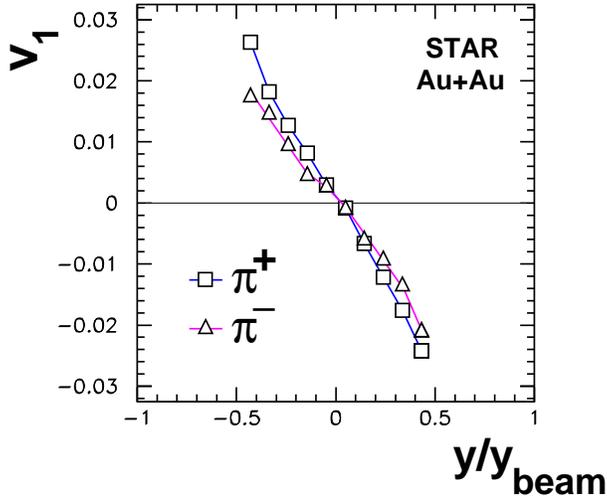}
   \caption{(Color online) Directed flow $v_1$ of positive and negative pions 
in intermediate centrality (10-40\%) Au+Au collisions at $\sqrt{s_{NN}}=7.7$~GeV. The data points in the figure are 
redrawn from~\cite{star2014}. The plotted error bars are statistical only, and remain below symbol size.}
 \label{fig:stardata}
\end{figure}

The results of the STAR Collaboration on charged pion directed flow in Au+Au collisions at 
$\sqrt{s_{NN}}=7.7$~GeV are shown in Fig.~\ref{fig:stardata}. The numerical values 
from~\cite{star2014} are redrawn as a function of the scaled pion rapidity, $y/y_\mathrm{beam}$, where 
$y_\mathrm{beam}$ is the rapidity of the incoming nucleus in the collision c.m. system. We 
decide on this scaled variable for an easier
 comparison with other collision 
energies, including our results 
 from~\cite{Rybicki-v1} 
 (see also discussions made 
in~\cite{na49prc,star-Pandit}). 
 As specified in~\cite{star2014}, the presented experimental data are obtained within a lower 
cut on pion transverse momentum, $p_T>0.2$~GeV/c, and an upper cut on pion total momentum, 
$p<1.6$~GeV/c. Three facts are immediately apparent from the figure:

\begin{enumerate}
 \item
 At this relatively low collision energy, the STAR data points cover a very appreciable 
region in $y/y_\mathrm{beam}$.
 \item
 Apart from the dominant and well-known trend of a smooth decrease with increasing rapidity, 
the data points clearly display a pronounced split of $v_1$ for $\pi^+$ and $\pi^-$.
 \item
 A minor deviation from
 antisymmetry about mid-rapidity is visible for the data points,
 in particular at the edges of the covered kinematical range.
 Following the detailed discussion 
 made in~\cite{star2014}, we regard 
 this simply as 
 an experimental 
uncertainty.
 \end{enumerate}

In order to get hold of electromagnetic effects on directed flow, we assume that the total 
values of positive and negative pion $v_1$ can be approximated by the sum of two terms:

\begin{equation}
v_1^{\pi^+} \approx {v_1^{flow}} + {v_1^{\pi^+,EM}}  \; ,
\label{eq:v1pi+}
\end{equation}
\begin{equation}
v_1^{\pi^-} \approx {v_1^{flow}} + {v_1^{\pi^-,EM}}  \; .
\label{eq:v1pi-}
 \end{equation}
 where the first, dominant, charge-independent term $v_1^{flow}$ corresponds to the total 
directed flow imposed by the strong interaction (e.g., by the hydrodynamical evolution of 
the system), while the second, smaller and charge-dependent term $v_1^{\pi^+,EM}$ 
($v_1^{\pi^-,EM}$) is induced by electromagnetic interactions. The approximate additivity 
postulated in 
 (\ref{eq:v1pi+}) and (\ref{eq:v1pi-}) was verified by Monte Carlo calculations,
somewhat simplified but
similar to these described in section~\ref{model}.

From our earlier studies~\cite{Rybicki-v1}, as well as from Monte Carlo simulations 
performed for the present work (section~\ref{results}), we know that in the range of 
rapidity considered here, the spectator-induced electromagnetic component of directed flow 
is, at least to a good approximation, opposite for opposite charges. Thus we postulate:

\begin{equation}
v_1^{\pi^+,EM} \approx -v_1^{\pi^-,EM} \; .
\label{eq:v1piem}
\end{equation}

By solving 
(\ref{eq:v1pi+})-(\ref{eq:v1piem}), we conclude that the electromagnetic 
 component 
of the directed flow presented in Fig.~\ref{fig:stardata} can be 
 obtained as:

\begin{equation}
v_1^{\pi^+,EM} \approx  \frac{1}{2} (v_1^{\pi^+} - v_1^{\pi^-})  \; ,
\label{eq:dv1pi+}
\end{equation}
\begin{equation}
v_1^{\pi^-,EM} \approx - \frac{1}{2} (v_1^{\pi^+} - v_1^{\pi^-}) \; .
\label{eq:dv1pi-}
\end{equation}

\begin{figure}[h]             
\centering
\includegraphics[width=0.5\textwidth]{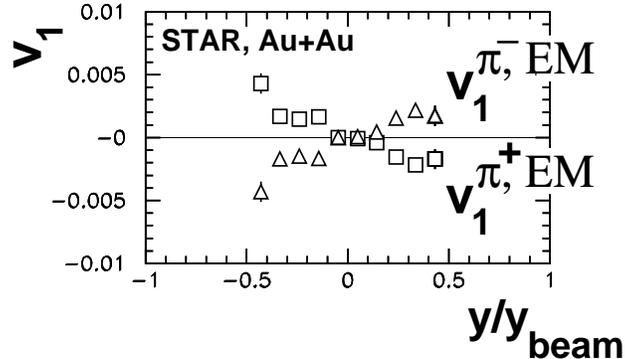}
   \caption{Electromagnetically-induced 
 directed flow for positive pions (squares) and for
negative pions (triangles), obtained from~(\ref{eq:dv1pi+})-(\ref{eq:dv1pi-}). 
The plotted error bars are statistical only.}
 \label{fig:stardataextr}
\end{figure}

This is shown in Fig.~\ref{fig:stardataextr}. The resulting values of $v_1^{\pi^+,EM}$
and $v_1^{\pi^-,EM}$ consistently reach up to about 0.002,
with the exception of one asymmetric outlier 
visible at the low edge of the covered 
kinematical range. This we regard as an experimental uncertainty as specified 
in point 3.~above.
 We conclude that the overall precision of the STAR data is sufficient to identify the - relatively 
small 
 - electromagnetic component of directed flow. As we will demonstrate in the subsequent 
parts of this paper, the magnitude of this electromagnetic component
 (Fig.~~\ref{fig:stardataextr}) provides information on the space-time evolution of the process 
of pion production in Au+Au collisions at central rapidities.


\section{Calculating the electromagnetically-induced directed flow}
\label{model}

We now turn to the description of our Monte Carlo method of estimating the part of charged pion 
directed flow induced by the electromagnetic interaction between final state pions and the two 
spectator systems. Our approach is essentially similar to that taken in our precedent 
works~\cite{Rybicki-v1,Rybicki-coulomb}, where a more detailed description can 
 be found. Only 
the aspects relevant for the present analysis will be discussed here.

\begin{figure}[t]             
\centering
\includegraphics[width=0.8\textwidth]{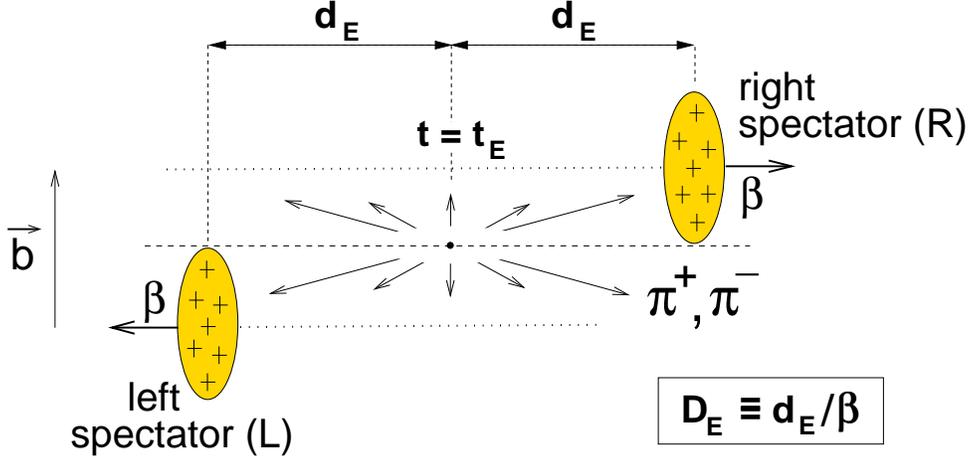}
   \caption{(Color online) Our simplified picture of the Au+Au collision at $\sqrt{s_{NN}}=7.7$~GeV.}
 \label{fig:model}
\end{figure}

Our aim is to provide a realistic estimate of spectator-induced electromagnetic effects on charged 
pion directed flow. At the same time, we wish to avoid a detailed discussion of the complex, poorly 
known mechanisms governing the dynamics of the Au+Au reaction.
 For this reason our approach is maximally 
simplified as explained below:

 \begin{enumerate}
 \item[(a)]
 We assume that the Au+Au collision takes place at a given impact parameter $b$, 
corresponding to the STAR sample of intermediate centrality reactions. This is illustrated 
in Fig.~\ref{fig:model}. The two spectator systems are approximated by homogeneous, 
Lorentz-contracted spheres, which follow their initial path with essentially unchanged 
velocities. The reaction plane is defined by the collision axis and by the impact 
parameter vector $\vec{b}$.
 \item[(b)]
 Charged pions ($\pi^+$, $\pi^-$) are assumed to be emitted from a single point in space 
(the original interaction point) and in a single moment in time ($t=t_E$). The resulting 
initial distance $d_E$ between the pion emission site and the two spectator systems is the 
unique free parameter of our simulation.
 For
 consistency with our precedent works this quantity will be expressed in terms of the
 reduced distance $D_E\equiv d_E/\beta$, where $\beta$ is the spectator
 velocity\footnote{For
 simplicity we set the velocity of light $c=1$ in the entire paper.}. 
 We note that
 at $\sqrt{s_{NN}}=7.7$~GeV the two quantities are equal within three percent.
 \item[(c)]
 The initial ($y$, $p_T$) distribution of the emitted pions (before the action of the electromagnetic 
field) is assumed to be similar to underlying nucleon-nucleon collisions and to obey wounded nucleon 
scaling~\cite{wnm}. Full azimuthal symmetry is assumed for these initially emitted pions.
 \item[(d)]
 The emitted charged pions are then numerically traced in the electromagnetic field induced by the 
spectator charges, until they reach a distance of 10,000 fm away from the original interaction point 
and from each of the two spectator systems. Spectator fragmentation is neglected. Effects induced by 
the participant charge, strong final state interactions, etc, are not considered.
 \end{enumerate}

Several clarifications must be added to points (a)-(d) above.

 \begin{enumerate}
 \item[(a)] {\em Geometry of the Au+Au collision}.
 The centrality of the STAR data~\cite{star2014} was defined by the number of charged 
particles emitted in the region of pseudorapidity $|\eta|<0.5$. For consistency, in the 
present work the number of participating nucleons $n_{part}$ was obtained by extrapolating 
the results of Glauber Monte Carlo calculations made with the same definition of centrality 
as above, in the range of $\sqrt{s_{NN}}$ from 20 to 200~GeV~\cite{Daugherity}. For 
intermediate centrality (10-40\%) Au+Au collisions, the extrapolation down to 
$\sqrt{s_{NN}}=7.7$~GeV gave the mean number of participants of about 166. The other 
geometrical characteristics of the collision have been estimated by means of a dedicated 
geometrical Monte Carlo simulation, discussed in detail in~\cite{Rybicki-coulomb}. This used 
the $^\mathrm{197}$Au Woods-Saxon density profile taken from~\cite{Daugherity} and assumed 
the elementary nucleon-nucleon cross-section equal to 30.6~mb, in agreement with existing 
data~\cite{pdg}. On that basis, the geometrical impact parameter corresponding to 166 
participants was found to be $b_{geom}=7.09$~fm. The center of gravity of each of the two 
spectator systems was found to be displaced by $\Delta b=2.00$~fm relative to that of the 
original Au nucleus. The average spectator charge was found to be $Q=45.7$~elementary 
units.\\
 After
 inspecting
 the spectator shape resulting from our 
 geometrical
 simulation, and 
considering its exact shape as unimportant for our present analysis, we modelled the two 
spectator systems as two homogeneously charged spheres. The sphere density was the standard 
nuclear density $\rho=0.17$/fm$^\mathrm{3}$ in the rest frame of each sphere. The center of 
each sphere was displaced by 2.00 fm in order to match the center of gravity of the spectator 
system. As a result, our effective impact parameter (distance of closest approach between the 
two sphere's centers) was $b=b_{geom}+2\Delta b=11.1$~fm.
 \item[(b)] {\em Pion emission}. As this will be shown in section~\ref{results}, our maximal 
simplification of initial conditions determining pion emission (i.e., the reduction of the 
emission zone to a single point in space and in time) gives a convenient way to estimate the 
sensitivity of the electromagnetic component of directed flow to the space-time evolution of 
the heavy ion reaction. In the present work, we will mostly focus on its sensitivity to 
the distance between the emitted pions and the spectator system as indicated in 
Fig.~\ref{fig:model}.

 \item[(c)] {\em Initial distribution of emitted pions}. We consider that the exact 
shape of the kinematical distribution of emitted pions has only a small influence on the 
electromagnetic component of $p_T$-integrated directed flow. On the other hand, the very 
nature of the spectator-induced electromagnetic interaction (acting on the pion $x$, $y$, 
$z$ trajectory over a sizeable period of time)
 implies
 the need for a reasonable 
parametrization of this distribution in terms of the complete 
 momentum vector defined, e.g., by 
 three variables~($y$, $p_T$, $\phi$). 
 As a result, we assume the pion ($y$, $p_T$) distribution to be similar to that in 
nucleon-nucleon events scaled up by the number of participant (wounded) nucleons. We 
describe it by means of an analytical parametrization of average pion 
$\left(\frac{\pi^++\pi^-}{2}\right)$ spectra in p+p collisions, obtained by the NA49 
experiment at $\sqrt{s_{NN}}=17.3$~GeV~\cite{pp}. A precise description of this 
parametrization is given in~\cite{Rybicki-coulomb}. The NA49 data are expressed in terms of 
the Feynman variable $x_F=2p_L/\sqrt{s}$ and of transverse momentum $p_T$, and cover the 
region from $x_F=0$ to $0.85$ and from $p_T=0$ to $2.1$~GeV/c in the collision c.m.s. This 
is more than sufficient for the present study, defined by the more restricted coverage of 
the experimental STAR data discussed in section~\ref{stardata}. For simplicity, 
distributions of positively and negatively charged pions are assumed to be identical. Feynman 
scaling in $x_F$~\cite{Feynman},
 and similarity 
 of $p_T$-spectra
 are 
 assumed from $\sqrt{s_{NN}}=17.3$~GeV down to 7.7~GeV.\\
 The aim of the present work being to estimate only the {\em electromagnetic 
component} of directed
 flow 
 (${v_1^{\pi^+,EM}}$ and 
 ${v_1^{\pi^-,EM}}$ from Eqs. 
 \ref{eq:v1pi+} and 
 \ref{eq:v1pi-}), the present simulation assumes {full azimuthal 
symmetry} in initial pion emission
 before the action of the electromagnetic field. As such, all the results presented in 
section~\ref{results} will correspond to directed flow induced exclusively by the 
electromagnetic interaction.


 \item[(d)] {\em Propagation of pions in the electromagnetic field}.
 Our numerical treatment of the motion of charged pions in the electromagnetic field induced 
by the two spectator systems was explained in detail in~\cite{Rybicki-coulomb}.
 A concise description can
 also
 be found in~\cite{Rybicki-v1}.
 Generally, the purely electrostatic fields of the two spectators in their respective rest 
frames (these we will denote as $\vec{E}'_{L}$ for the ``left'' spectator and 
$\vec{E}''_{R}$ for the ``right'' spectator from Fig.~\ref{fig:model}) are transformed to 
the overall collision c.m.s. This results in the emergence of both electric and magnetic 
fields. For the ``left'' spectator one writes:
 \begin{equation}
\begin{split}
\vec{E}_{L} &= \gamma \vec{E}'_{L}
 - \frac{\gamma^2}{\gamma+1}
\; {\vec{\beta}_{L}} \; \left( {\vec{\beta}_{L}} \cdot 
\vec{E}'_{L}
\right)
\; ,
 \\
\vec{B}_{L} &= \gamma \left( {\vec{\beta}_{L}} \times 
\vec{E}'_{L} 
\right)
\; .
\end{split}
\label{transformation}
\end{equation}
 In the equations above,
 ${\vec{\beta}_{L}}$ is the vector of velocity of the ``left'' spectator 
($|{\vec{\beta}_{L}}|=\beta$ in Fig.~\ref{fig:model}) and
 $\gamma=(1-\beta^2)^{-1/2}$. Analogous equations can be written for the ``right'' 
spectator, yielding the two corresponding fields $\vec{E}_{R}$ and $\vec{B}_{R}$ in the 
nucleus-nucleus
collision c.m.s. The Lorentz force $\vec{F}_{\pi}$ acting on the pion results from the 
combined action of electric and magnetic fields:
\begin{equation}
\frac{d \vec{p}_{\pi}}{dt} = \vec{F}_{\pi} =
 q_{\pi} \left( \vec{E} 
+ {\vec{\beta}_{\pi}} \times \vec{B} \right) \; ,
\label{Lorentz_force}
\end{equation}
 where ${\vec{p}_{\pi}}$ and ${\vec{\beta}_{\pi}}$ are the pion momentum and velocity vectors,
$q_\pi$ is the pion charge, while
$ \vec{E} = \vec{E}_{L} + \vec{E}_{R} $ and 
$ \vec{B} = \vec{B}_{L} + \vec{B}_{R} $ are standard 
 superpositions of fields from the two sources.
The pion trajectory $\vec{r}_{\pi}(t)$ is given by the classical relativistic equation of motion:
\begin{equation}
\frac{d \vec{r}_{\pi}}{d t} = \vec{\beta}_{\pi} = 
\frac{\vec{p}_\pi}{\sqrt{p^2_\pi + m^2_\pi}} \; ,
\label{velocity}
\end{equation}
 where $m_{\pi}$ is the pion mass, and the pion momentum ${\vec{p}_{\pi}}$ is obtained from
Eq.~(\ref{Lorentz_force}).
 Our approach explained above takes account of relativistic effects, including in particular 
also retardation~\cite{Jackson}. Technically, the propagation of the charged pion in the 
electromagnetic field is performed numerically, by means of an iterative procedure made in 
small steps in time. This is done with variable step size which depends on the actual 
distance of the pion from the nearest spectator system. The iteration proceeds until the 
pion is at least 10,000 fm away from the interaction point and from each of the two 
spectator systems in their respective rest frames. We note that negative pions, which do not 
escape from the potential well induced by the spectator system are rejected and do not enter
 into
 our final state distributions.

 \end{enumerate}


\section{Results}
\label{results}

This section contains the discussion of results of our simulation of 
electromagnetically-induced directed flow, caused by the electromagnetic 
interaction between the spectator systems and charged final state pions. These results will 
be compared to the charge-dependent part of 
the pion
directed flow 
 measured
 by the STAR experiment.

All the Monte Carlo results presented here have been obtained for intermediate centrality 
Au+Au collisions at $\sqrt{s_{NN}}=7.7$~GeV, following the procedure described in 
section~\ref{model}. They are integrated over transverse momentum $p_T$ in the range from 
0.2 to 1.6~GeV/c. It has been verified that these results will practically not change 
when applying the cut on total momentum $p<1.6$~GeV/c actually used for the STAR data 
(section~\ref{stardata}) instead of the upper cut on $p_T$. Thus the results of the 
simulation are directly comparable to the values extracted from experimental data.

 We remind that as it is described in section~\ref{model}, the 
presented Monte Carlo 
 simulations
 correspond to effects resulting
 {\em exclusively from the electromagnetic interaction}. On purpose, the part of 
pion
directed flow resulting from the strong interaction (denoted $v_1^{flow}$ in 
section~\ref{stardata}) is not included in the 
 simulation.
 The same should be valid for the 
charge-dependent part of experimentally observed directed flow, which we
 extracted from
the
STAR data in section~\ref{stardata}, Fig.~\ref{fig:stardataextr}.

 Fig.~\ref{fig:allde} shows the electromagnetically-induced directed flow of
positive and negative pions (denoted 
$v_1^{\pi^+,EM}$ and $v_1^{\pi^-,EM}$ as in Eqs.~\ref{eq:v1pi+} and~\ref{eq:v1pi-}),
 simulated assuming six different values of the reduced distance 
between the pion formation zone and the spectator system: $D_E=0$, 0.5, 1, 2, 3, and 5 fm. Several observations can be made from the Figure:

\begin{figure}[t]             
\centering
\includegraphics[width=0.95\textwidth]{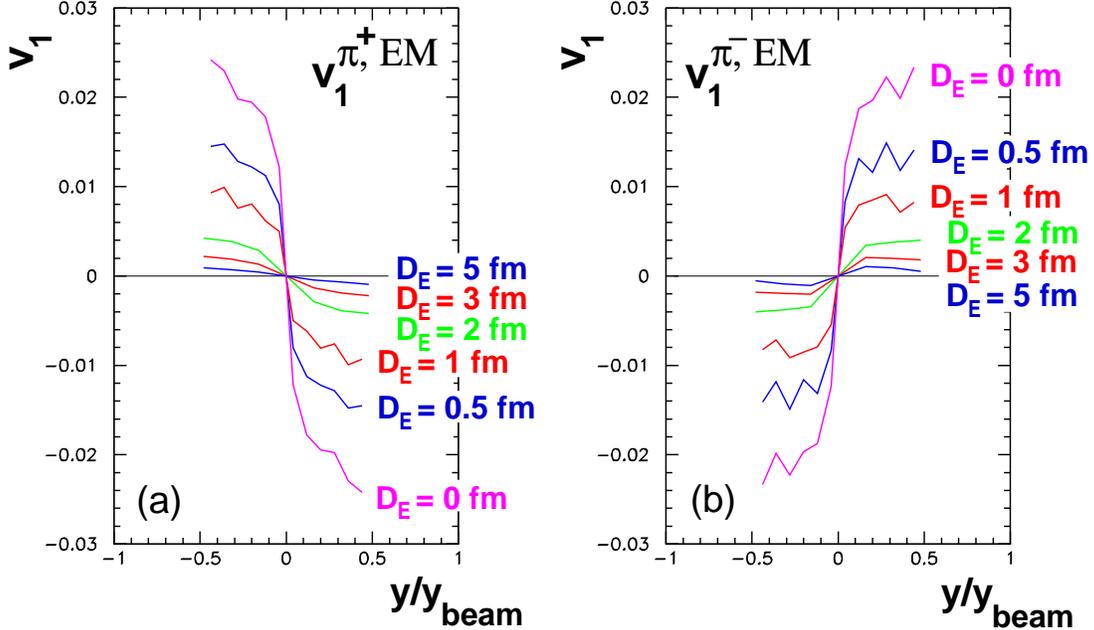}
   \caption{(Color online) Electromagnetically-induced directed flow of (a) positive and 
(b) negative pions in intermediate centrality Au+Au collisions at $\sqrt{s_{NN}}=7.7$~GeV. 
The six curves in each panel correspond to six different values of the reduced distance 
$D_E$ assumed in the simulation.}
 \label{fig:allde}
\end{figure}

\begin{figure}[t]             
\centering
\includegraphics[width=0.95\textwidth]{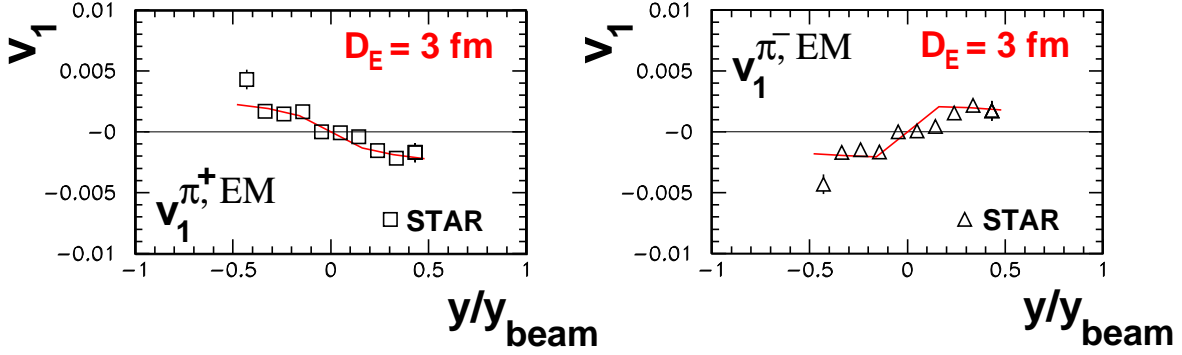}
   \caption{(Color online) Electromagnetically-induced directed flow of positive pions 
(left panel) and negative pions (right panel) in intermediate centrality Au+Au collisions 
at $\sqrt{s_{NN}}=7.7$~GeV. The result of our Monte Carlo simulation made assuming 
$D_E=3$~fm (red curve, same as in Fig.~\ref{fig:allde}), is compared to 
values 
extracted from the experimental STAR data~\cite{star2014} 
(data points, same as in Fig.~\ref{fig:stardataextr}).}
 \label{fig:datamodel}
\end{figure}

 \begin{enumerate}
 \item
 The electromagnetic interaction between final state pions and spectators can result in 
absolute values of directed flow reaching maximally up to 0.025 in the considered range of rapidity. 
This is far lower than what we obtained for pions close to beam rapidity, where our 
prediction for electromagnetically-induced directed flow~\cite{Rybicki-v1} exceeded
absolute values of 0.2, in good agreement with data obtained by the WA98 experiment at the 
CERN SPS~\cite{wa98}.
 \item
 Nevertheless, the values 
of $v_1^{\pi^+,EM}$ and $v_1^{\pi^-,EM}$
 shown
 in Fig.~\ref{fig:allde}
 have to be considered as potentially very sizeable 
if compared to 
the
{\em total} values of pion directed flow observed by STAR, Fig.~\ref{fig:stardata}. In 
the extreme
 case of ``immediate'' pion production ($D_E=0$~fm), the 
obtained values of $v_1^{\pi^+,EM}$ would indeed become {comparable} to {total} values of $v_1$ 
measured by the experiment. 
This shows that the 
 overall
magnitude of the spectator-induced 
electromagnetic interaction is large enough to exert 
potentially 
very important,
charge-dependent effects on pion directed flow.
 \item
 At the same time, the electromagnetically-induced directed flow displays a very strong 
sensitivity to the postulated distance between the pion emission site and the two spectator 
systems. A change in $D_E$ of the order of 1~fm results in perfectly visible 
changes in
$v_1^{\pi^+,EM}$ and $v_1^{\pi^-,EM}$; these
 changes
 can exceed 0.01 in the considered range of 
rapidity. This implies that {\em the spectator-induced electromagnetic effect on pion directed 
flow brings new, independent information on the space-time evolution of the system created in the Au+Au 
collision}, and in particular on the way the emission of final state pions evolves in position space and in time.
 \item
 Specifically, the experimental data from the STAR detector have a sufficient 
discriminative power to put
 significant constraints on the postulated space-time 
scenario of pion emission.
 This is evident in Fig.~\ref{fig:datamodel}, where 
the charge-dependent component of the directed flow measured by the STAR
Collaboration 
clearly appears to favour the value of $D_E=3$~fm as the best fit to the data.
 \end{enumerate}
%
%
 We conclude from the above that the 
 magnitude of the
 electromagnetic interaction between final state pions 
and
 the
 charged spectator systems is largely sufficient to explain the effect of charge 
splitting present in the STAR data from~\cite{star2014}. We also find that a 
scrutiny of these data in view of the above electromagnetic effect can bring new insight 
into the space-time evolution of 
 pion production in 
 the nucleus-nucleus reaction.
 This can constitute a new source of information on the 
 space-time 
 properties 
of the system created in 
the heavy ion collision, completely independent 
 from other sources
 such as
 pion interferometry.

\section{Implications}
\label{implications}

 Here 
 we will discuss some of the 
implications 
 of 
 the 
 observations made in section~\ref{results}.
 The 
 discussion will address the results of our present analysis, but will also include 
some  of the findings made in
our earlier studies 
of spectator-induced electromagnetic effects on pion directed flow~\cite{Rybicki-v1} and on 
$\pi^+/\pi^-$ ratios~\cite{Klusek2012} in peripheral Pb+Pb collisions at
 the top SPS energy,~$\sqrt{s_{NN}}=17.3$~GeV. 
 We will also address
some of the issues 
 discussed
 in Ref.~\cite{kharzeev2014}.

\subsection{Pion emission distance as a function of pion rapidity}
\label{rap-dep}

As it was said in the precedent sections, the unique free parameter in our 
  Monte Carlo
 simulation 
is the pion emission distance ($d_E$ in Fig.~\ref{fig:model}), namely, the
distance between the pion formation zone and the two spectator 
systems. The comparison of the
results of our simulation to electromagnetic effects seen in experimental data (as performed in section~\ref{results}, Fig.~\ref{fig:datamodel} above) allows us to define the {\em optimal} value of the parameter $d_E$, most favoured by the experiment. In spite of the simplicity and of the somewhat effective character of our Monte Carlo study, it is nevertheless interesting to consider how this optimal pion emission distance 
varies as a function of pion rapidity.

 This is made in Table~\ref{de}, where we summarize our findings on the optimal pion emission distance $d_E$ as they result 
from the present work and from our earlier studies. 
 Altogether three experimental data sets are considered in the table, see respectively~\cite{star2014},~\cite{Rybicki-epshep}, 
and~\cite{wa98}. 

 We note that for the present analysis, $d_E\equiv\beta D_E\approx D_E$ as discussed in 
section~\ref{model}.
 This gives us $d_E\approx 3$~fm, as evident from the comparison with the STAR 
data
 made in section~\ref{results}.
The comparison of our simulation to the data 
set on Pb+Pb collisions~\cite{Rybicki-epshep} yields the optimal value of $d_E$ in the range between 0.5 and 
1~fm, while the comparison to the data set~\cite{wa98}
 suggests
 $d_E$ in the range from 0 to 1~fm. 
 A more detailed discussion of the two latter comparisons can be found in 
Refs.~\cite{Klusek2012} and~\cite{Rybicki-v1}, respectively.

\begin{table}[tb]
\begin{center}
\begin{tabular}{  c  c  c  c  l  c  }
 Ref. & reaction    &  pion rapidity    & ~~~~observable~~~~ & ~exp.~data~ & ~resulting~$d_E$~  \\
\hline
~this work~          & ~Au+Au~ &  ~{\small $-0.5<y/y_\mathrm{beam}<0.5$}~ & ~directed flow~ & ~STAR~\cite{star2014}~ & ~$\approx 3$~fm~ \\
~\cite{Klusek2012}~  & ~Pb+Pb~ &  ~$~0.64<y/y_\mathrm{beam}<1.1$~ & ~$\pi^+/\pi^-$~ratio~ & ~NA49~\cite{Rybicki-epshep}~ & ~$0.5-1$~fm~ \\
~\cite{Rybicki-v1}~  & ~Pb+Pb~ &  ~$~~0.9<y/y_\mathrm{beam}<1.3$~ & ~directed flow~ & ~WA98~\cite{wa98}~ & ~$0-1$~fm~ \\
\end{tabular}
\end{center}
 \caption{Summary of our findings on the distance $d_E$ from the present work in comparison 
to our earlier studies.
 The Au+Au collisions are 
taken
at $\sqrt{s_{NN}}=7.7$~GeV and have intermediate centrality, while the Pb+Pb reactions at $\sqrt{s_{NN}}=17.3$~GeV are peripheral. For Ref.~\cite{Klusek2012}, the range in rapidity corresponds to
$0.1<x_F<0.4$ at the average considered value of $p_T$.}
\label{de}
 \end{table}

\begin{figure}[htb]             
\centering
\includegraphics[width=0.55\textwidth]{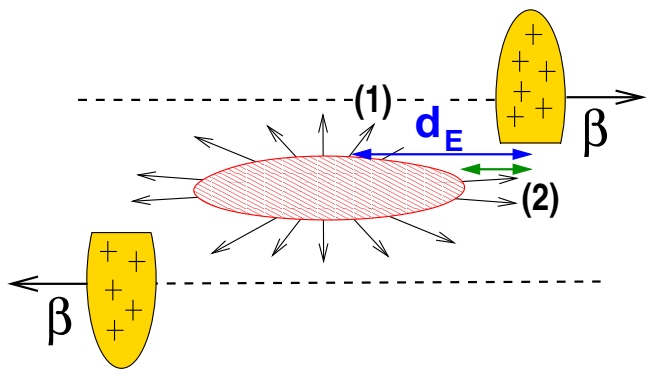}
   \caption{(Color online) Schematic picture of the
 longitudinally
 expanding system created in the heavy ion 
collision. The distance to the spectator system ($d_E$) is larger for pions emitted at lower
rapidity (1) than for pions emitted at higher rapidity (2).}
 \label{fig:expanding}
\end{figure}

 In spite of complications arising from 
the 
difference in collision energy and centrality between the present work and 
the two other analyses, a consistent trend 
is apparent
in Table~\ref{de}. At higher values of $y/y_\mathrm{beam}$, and independently on the considered observable, pions appear to be emitted {relatively {close} to the spectator system}, with $d_E$ remaining below 1~fm. At more central rapidities the pion emission distance {increases significantly} ($d_E\approx 3$~fm). This is to be expected for 
the 
{\em 
 longitudinally
 expanding system} created in the heavy ion collision, where pions at higher rapidity will
decouple 
closer to the spectator as shown in Fig.~\ref{fig:expanding}.


The above observations bring, in our view, important consequences. For 
the first time, information on the space-time characteristics of this 
 longitudinally
 expanding, strongly-interacting system responsible for particle 
production is being provided {\em by means of the spectator-induced 
electromagnetic interaction}. This information remains {\em completely 
{independent}} from that provided by femtoscopic 
analyses~\cite{alice-hbt} or combined blast-wave model 
fits~\cite{na49-hbt-blastwave}.
 Also,
as 
the latter electromagnetic interaction is, by itself, model-independent, this gives
 hope for a 
reduction of uncertainties currently present in our knowledge of the expanding 
matter created in heavy ion collisions, and for a verification of existing phenomenological 
models of non-perturbative pion production.

As it is evident from Table~\ref{de}, information provided by the spectator-induced electromagnetic interaction is not confined to any specific region of pion rapidity. On the contrary, it can be used anywhere down from the central (``mid-rapidity'') region up to and beyond beam rapidity. 
While our present analysis still remains simplified and 
quite
rudimentary,
already at the present moment
it seems 
clear 
that more detailed simulations can bring new insight into the 
complete 
$(x,y,z,t)$ distribution of the pion emission zone drawn in Fig.~\ref{fig:expanding}.


\subsection{Comments on other possible effects}
\label{com}

While the aim of this paper is to clarify the role of the electromagnetic interaction between spectators and final state pions in the charge splitting of pion directed flow, some other possible electromagnetic effects should also be commented upon. A first, evident candidate is the electromagnetic interaction induced by the participant charge rather than the spectator charge. For intermediate centrality Au+Au collisions measured by the STAR Collaboration which constitute the basis for the analysis made here, it cannot be {\em a priori} excluded that the considerable net participant charge will also exert some influence on the observed charge splitting of directed flow. This effect, neglected in our analysis, should in principle be taken into account in future, more detailed studies. On the other hand, we note that the issue of participant charge was studied and commented upon in Ref.~\cite{kharzeev2014}, in the context of magnetically-induced electric currents acting in the QGP and of their influence on the charge splitting of directed flow. Here, the authors found that the role of participant charge was small (maximally~10\%) w.r.t. that induced by the spectator charge.

 The work~\cite{kharzeev2014} cited above 
 touches another, very interesting problem which necessitates a comment in view of the results obtained in the present analysis. Studying the electric currents induced in the QGP, the authors conclude that the latter result in a charge-dependent directed flow. The results presented in Ref.~\cite{kharzeev2014} for positive (negative) pions show this directed flow as mostly negative (positive) at positive rapidity, which is qualitatively similar to our results presented in Figs~\ref{fig:allde} and~\ref{fig:datamodel}. The magnitude of the effect strongly depends on pion transverse momentum; limiting our considerations to the range discussed in the present paper, $0<p_T<1.6$~GeV/c, the curves presented in~\cite{kharzeev2014} reach maximal absolute values of $v_1\approx 0.00004$ at $p_T=1$~GeV/c for Pb+Pb collisions at the LHC energy $\sqrt{s_{NN}}=2.76$~TeV, and of $v_1\approx 0.00012$ at $p_T=1$~GeV/c for Au+Au collisions at top RHIC energy, $\sqrt{s_{NN}}=200$~GeV (both collisions are considered at 20-30\% centrality). This is in principle well below the magnitude of charge-dependent directed flow 
 which we obtained from the STAR data at $\sqrt{s_{NN}}=7.7$~GeV (up to 
about 0.002, see section~\ref{stardata}, Fig.~\ref{fig:stardataextr}). 
However, the authors of~\cite{kharzeev2014} expect the effects of 
magnetic fields to increase with decreasing collision energy, and point at the low 
energy STAR data as showing hints of the phenomena that they had 
described.

These constatations call, in our view, for a
 detailed scrutiny in the context of our results obtained in section~\ref{results}. Indeed, our work suggests that the electromagnetic interaction between spectators and {final state} pions is, {alone}, 
 sufficient to explain the charge splitting of directed flow present in 
the
STAR data~\cite{star2014}, with no apparent necessity to involve 
phenomena related to the electric conductivity of the quark-gluon 
plasma. This was demonstrated in Fig.~\ref{fig:allde} and most of all in 
Fig.~\ref{fig:datamodel}. As we take our electromagnetic effect on 
final state pions as a straight-forward and {\em unavoidable} consequence of the 
presence of spectator charges in the collision, it seems to us that 
 this effect
 must be 
 considered
 as a baseline
 whenever
 postulating
 any more sophisticated phenomenon as the one discussed in~\cite{kharzeev2014}.

More
 detailed
 studies, involving in particular also the energy dependence of our 
effect on final state pions, are necessary in order to get more insight 
into the possible interplay between these two effects as a function of 
collision centrality and energy\footnote{Similarly to phenomena described in~\cite{kharzeev2014}, we also expect our effect to increase with decreasing energy.}.


\section{Summary and conclusions}
\label{conclusions}

To the best of our knowledge, the present work was the first analysis of 
the role played by the spectator-induced electromagnetic interaction in 
building up the directed flow of charged final state pions produced at 
central rapidities. This analysis was limited to the spectator-final 
state pion EM interaction and did not include more sophisticated 
phenomena related to the electrical conductivity of the quark-gluon 
plasma. On the basis of our numerical simulation, and assuming 
intermediate centrality Au+Au collisions at $\sqrt{s_{NN}}=7.7$~GeV, we 
conclude that the spectator-induced electromagnetic interaction can 
result in a charge splitting of $\pi^+$ and $\pi^-$ directed flow which will come on top of phenomena resulting from the strong interactions (Eqs.~\ref{eq:v1pi+},~\ref{eq:v1pi-}).

The magnitude of this charge splitting appears to strongly depend on the 
actual distance between the pion emission site (pion at freeze-out) and 
the spectator system. As such, this effect brings new, independent 
information on the space-time evolution of pion production.

In the extreme case of ``immediate'' pion production, the 
spectator-induced electromagnetic interaction would induce a maximal 
charge splitting 
of $v_1^{\pi^+}$ 
and $v_1^{\pi^-}$
which could reach up to $2\cdot 0.025=0.05$ 
in the considered range of rapidity 
(as apparent in Fig.~\ref{fig:allde}). The charge-dependent component of directed flow, 
extracted from the experimental STAR data~\cite{star2014} and fairly well 
described by our simulation, suggests a much lower value 
for this charge splitting,
up to about 
$2\cdot 0.002=0.004$ (as apparent in Fig.~\ref{fig:datamodel}).

Adjusting our simulation to fit the experimental STAR data, and from the 
comparison of the present analysis to our earlier studies made for pions 
at higher rapidity, we formulate conclusions on the evolution of the 
pion emission distance as a function of rapidity. The distance between 
the pion emission at freeze-out and the spectator system appears to 
decrease with increasing pion rapidity, reflecting the longitudinal 
expansion of the system.  Thus for the first time, information on the 
space-time characteristics of the strongly-interacting 
system 
 created in the collision
 is being provided {by 
 means of
 the spectator-induced electromagnetic interaction}.

Finally, we comment on the effects of electric currents induced in the 
QGP and studied in Ref.~\cite{kharzeev2014}. The magnitude of our EM 
effect on final state pions being largely sufficient to describe the 
charge splitting of directed flow apparent in the STAR data~\cite{star2014}, 
we think that this effect should be taken as a baseline 
whenever considering 
any more sophisticated phenomenon like the one discussed 
in~\cite{kharzeev2014}. Further, more advanced studies are needed in 
order to differentiate between these two effects.\\

{\bf Acknowledgments}\\

The authors warmly thank Yadav Pandit and the STAR Collaboration for 
providing the numerical values for the published STAR data.
 This work was supported by the Polish National Science Centre 
(on the basis of decision no. DEC-2011/03/B/ST2/02634).

 \end{document}